\documentstyle[aps,prl,twocolumn]{revtex}

\input psfig

\begin{document}
\draft
\preprint{DART-HEP-98/03}
\title{Gravitational Waves from Collapsing Vacuum Domains}

\author{Marcelo Gleiser\footnote{NSF Presidential Faculty Fellow. 
email: gleiser@dartmouth.edu.} 
and Ronald Roberts\footnote{email:
ronald.roberts@dartmouth.edu.} }

\address{Department of Physics and Astronomy, Dartmouth College,
Hanover, NH 03755, USA}

\date {DART-HEP-98/03~~\today}

\maketitle

\begin{abstract}
The breaking of an approximate discrete symmetry, the final stages of a
first order phase transition, or a post-inflationary biased probability
distribution for scalar fields are possible cosmological scenarios 
characterized by the presence of unstable domain wall networks. Combining 
analytical and numerical techniques, we show that the non-spherical collapse of 
these domains can be a powerful source of gravitational waves.
We compute their contribution to the stochastic background of
gravitational radiation and explore their observability by
present and future gravitational wave detectors.
\end{abstract}

\pacs{04.30.Db, 98.80.Cq}

In order to further our understanding of the physical processes that took
place early in the history of the Universe we should explore
potential events which may have left an imprint detectable by
current or future experiments. One exciting possibility is that
certain primordial processes generated
a stochastic background of gravitational waves. During the past decade or so,
several cosmological sources (as distinguished from astrophysical sources such 
as coalescing binary systems \cite{EANNA})
of stochastic gravitational waves have been
proposed \cite{ALLEN}. These include inflationary models \cite{INF},
cosmic strings \cite{COS-STRING}, strong first order phase transitions 
\cite{FIRST1,FIRST2}, and nontopological solitons \cite{NTS}. 

All ground-based
detectors (interferometric or resonant bar), such as LIGO or 
VIRGO, probe the frequency interval
$10{\rm Hz}\lesssim f \lesssim 10^4{\rm Hz}$. Lower frequencies are to be
probed in space, as with the planned 
LISA mission, which has a projected sensitivity of 
$10^{-4}{\rm Hz}\lesssim f\lesssim 1{\rm Hz}$.

Here we propose another powerful source of primordial gravitational waves,
the non-spherical collapse of bounded domain walls, or bags, which separate
different vacuum regions. 
Zel'dovich {\it et al.}, have shown that the
appearance of domain walls is a direct consequence of the 
breaking of
a discrete symmetry \cite{ZEL}. In the 
same work, domain walls were shown to be incompatible with big-bang
cosmology, as they would
create a power law expansion ruled-out by
observations. Thus, if domain walls were created they had to disappear.
One of the mechanisms proposed by Zel'dovich {\it et al.} was to consider an
{\it approximate} discrete symmetry as opposed to an exact one; the difference
in energy density between the two vacua generates a pressure force that 
eliminates
the walls. Other scenarios for the disappearance of
the domain walls were investigated in Ref. \cite{GGK}.

The existence of walls and other topologically stable defects were explored by
Kibble, who also suggested the possibility of an approximate
symmetry as a mechanism to eliminate the walls \cite{KIBBLE}. Several authors
explored this idea further
\cite{APPROX,GGK}. Recently, it was shown that unstable
domain walls could also be created during a post-inflationary non-thermal
phase transition \cite{LALAK}. The evolution of the
unstable domain-wall network has been studied numerically \cite{SARKAR}
and analytically \cite{MARK}. 

We can envisage at least three scenarios where unstable domain wall
networks can be generated: i) the breaking of an approximate discrete symmetry
in a thermal phase transition; ii) biased fluctuations in a non-thermal 
post-inflationary scenario; iii) during the final stages of a first order phase
transition, as bubbles of the true vacuum percolate, leaving a disconnected 
network of shrinking false vacuum domains.

We can model all these scenarios with 
a biased double-well potential for a real scalar field,

\begin{equation}
\label{pot}
V(\phi) = {{\lambda}\over 4}\left (\phi^2 -\phi_0^2\right )^2 +
h \phi_0\phi\left ( {1\over 3}\phi^2 - \phi_0^2\right )~,\
\end{equation}
where $h>0$ is a dimensionless
constant which biases the potential towards the $+\phi_0$ minimum.
The energy density difference between the two minima is $\Lambda =
{4\over 3}h\phi_0^4$, while the energy density barrier between the maximum
at $\phi_0=-(h/\lambda)\phi_0$ and the global minimum ($+\phi_0$) is
$B= {{\lambda }\over 4}\phi_0^4[1-{\cal O}(h/\lambda)]$.
Coordinates scale as ${\tilde x} = \sqrt{\lambda}\phi_0~x$,
while energies scale as ${\tilde E} =  (\sqrt{\lambda}/\phi_0)~E$.

The formation of a domain wall network during a phase transition has been 
discussed
in several works \cite{KIBBLE,COS-STRING,GGK,LALAK}. There are two
possibilities, determined by the probability the field has of landing on either
vacuum, $p_{\pm}$. There 
is a critical ``percolation probability'' for a given vacuum, $p_c$. For cubic
lattices, $p_c=0.31$, the value we will adopt here \cite{STAUFFER}. If 
both $p_+$ and $p_-$ are larger than $p_c$ both vacua percolate,
being separated by a domain wall stretching across the lattice. At formation, 
the 
wall will be rather convoluted, with local average curvature given by the
typical fluctuation scale, the correlation length $\xi$. If only $p_+>p_c$,
most of the volume will be in the $+$-vacuum, with isolated clusters (bags)
of the negative
vacuum with a distribution function, $f(r)\sim r^{-1.5}\exp [-r]$, where $r$ is
the number of cells of unit volume $V_{\xi}\simeq \xi^3$ in a given cluster
\cite{STAUFFER}. We
will call this case the ``no-percolation'' case.

The energy density of an isolated domain is given by

\begin{equation}
\label{energy}
\rho[\phi_{\rm b}] = {1\over 2}\left ({{\partial \phi_b}\over 
{\partial t}}
\right )^2 + {1\over 2}\nabla\phi_b\cdot \nabla\phi_b + V(\phi_b)~~,
\end{equation}
where $\phi_b({\bf x},t)$ is the field configuration describing the bag at time
$t$. We define the average radius
of an isolated domain by 
$R_{\rm av} = \left (\int dV~ \mid {\bf x}\mid ~\rho[\phi_{\rm b}]\right )/
\int dV~ \rho[\phi_{\rm b}]$. 

Two forces will act on the walls; the
tension, $P_T$, which will act to straighten the walls, and the vacuum pressure,
$P_{\Lambda}$, which will cause the domains to shrink. 
Here, we will consider 
two possibilities. First, the ``no-percolation case'', where the false 
vacuum bags
disappear very quickly after they form, that is, with $P_{\Lambda}> P_T$ right
at formation. This case is characterized by having several small bags within the
horizon.
Second, we will consider the ``percolation case'', where the 
tension force will straighten the walls for a while, up to the horizon
scale, before the vacuum pressure acts to accelerate the walls against each 
other,
causing the formation of large unstable bags \cite{GGK}.

We compute the output in gravitational radiation from the collapsing domains
within the full, linearized gravity approximation \cite{WEINBERG}. 
More details will be provided in a forthcoming publication. 
In the meantime, the reader may consult 
Ref. \cite{FIRST2}, as we follow a similar approach.

The computation of the spectrum in gravitational waves from collapsing 
3-dimensional domains in the linearized gravity approximation is a rather
cumbersome and computer-intensive
task. As the authors of Ref. \cite{FIRST2}, we have limited ourselves here
to studying the spectrum for situations with axial symmetry. In our case,
as toy models to more realistic bag configurations, 
we considered two types of domains, ellipsoids and deformed Gaussians.
Contrasting our results with the full 3-dimensional calculations for the 
simpler quadrupole approximation, we conclude
that here we are providing a lower bound for the total output in gravitational
radiation.

The total energy radiated in the direction ${\bf {\hat k}}$ into the solid
angle $\Omega$ at frequency $\omega$ is given by,

\begin{eqnarray}
\label{g_energy}
{{dE_{\rm GW}}\over {d\omega d\Omega}} & = & G\omega^2\big [T^{zz}({\bf {\hat k}},
\omega)\sin^2\theta +
T^{xx}({\bf {\hat k}},\omega)\cos^2\theta  \nonumber \\
& - & T^{yy}
({\bf {\hat k}},\omega) -2T^{xz}({\bf {\hat k}},\omega)\sin\theta\cos\theta
\big ]^2
\end{eqnarray}
where $T^{ij}({\bf {\hat k}},\omega)$ are the Fourier-transformed 
spatial components of the energy
momentum tensor for the field configuration $\phi_b$ \cite{FIRST2}.

From simple scaling arguments, the integrated energy radiated in gravitational
waves can be written as,
\begin{equation}
\label{e_total}
E_{\rm GW}(R_{\rm av}) = \epsilon(R_{\rm av})G\sigma^2R_{\rm av}^3~,
\end{equation}
where $\sigma$ is the surface energy density for the domains, and
$\epsilon(R_{\rm av})\equiv \epsilon_0\left ({{R_{\rm av}}\over {\xi_0}}
\right )^{\alpha}$ is the efficiency parameter, with $\epsilon_0$ 
a number to be determined. $\xi_0$ is the $T=0$ correlation length.

For ellipsoids, we wrote

\begin{equation}
\label{ellip}
\phi_b(r,z) = \phi_+ +{{(\phi_- - \phi_+)}\over 2}\left (
1- \tanh (\varepsilon/\sqrt{2})\right )~~,
\end{equation}
where $\varepsilon$ is the solution of $r^2/(a+\varepsilon)^2 +
z^2/(b+\varepsilon)^2 = 1$. For deformed Gaussians,
we wrote,

\begin{equation}
\label{gaussian}
\phi_b(r,\theta) = \phi_+ + (\phi_- - \phi_+){\rm exp} 
\left [- {{r^2}\over {\left [R_0
\left (1+\delta R\right )\right ]^2}}\right ]~,
\end{equation}
where $\delta R = \sum_{l,m}R_{l,m}Y_{l,m}(\theta,\varphi)$ measures the 
distortion from spherical symmetry. For axial symmetry we kept $m=0$.
These
configurations are then used as initial data for solving numerically the
Klein-Gordon equation for $\phi(r,z,t)$. As the configuration evolves in time,
we compute the emission of gravitational
radiation using Eq. \ref{g_energy}.

As shown in Fig. 1, the energy of the initial configurations can be well
fitted by $E_0(R_{\rm av}) = E_0^0\left ({{R_{\rm av}}\over {\xi_0}}
\right )^{\gamma}$. 
{}For spherically symmetric domains with degenerate vacua $\gamma=2$.
Writing also $E_0(R_{\rm av})=\sigma R_{\rm av}^2$, we obtain that the surface
tension scales as $\sigma={{E_0^0}\over {\xi^2_0}}\left (
{{R_{\rm av}}\over {\xi_0}}\right )^{\gamma-2}$.
From left to right, the sample points shown are: for ellipsoids
$(a~ b)= [(3~ 1.5), (2~ 4), (4~ 2), (4~ 6), (6~ 4), (6~ 10), (10~ 6)]$;
for deformed
Gaussians $(\ell~ R_0 ~ \delta R) $ $= [
(4~ 3 ~.5),  (3~ 3 ~.5), (4~ 4 ~.8), (3~ 5 ~.3), (4 ~5 ~.5), (4 ~6 ~.8),
(4 ~8 ~.5)]$.

\begin{figure}
\hspace{2.3in}
\psfig{figure=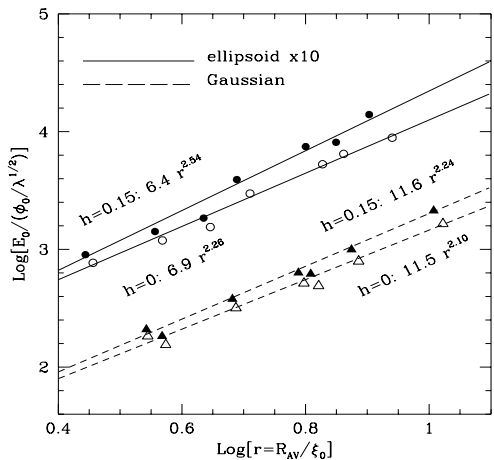,width=2.7in,height=2.5in}
\caption{The initial energy of the nonspherical domains vs. their
average radius for degenerate ($h=0$) and nondegenerate ($h=0.15$)
potentials. 
The power law fits are shown explicitly.}
\end{figure}

In Fig. 2, we show the integrated energy in gravitational waves, which
can be fitted by a simple power law relation, [$M_{Pl}=1.2\times 10^{19}$ GeV
is the Planck mass]

\begin{equation}
\label{e_total_fit}
E_{\rm GW}(R_{\rm av}) = E_{\rm GW}^0\left ({\phi_0\over {M_{Pl}}}\right )^2
\left ({{R_{\rm av}}\over {\xi_0}}\right )^{\beta}~~.
\end{equation}

Comparing Eqs. \ref{e_total} and \ref{e_total_fit}, we obtain that
$\beta = \alpha +2 \gamma  - 1$, and 
$\epsilon_0 = {{{\tilde E}_{\rm GW}^0}\over {\sqrt{2}
({\tilde E}_0^0)^2}}[2\sqrt{\lambda}]^{-1}$. The quantities with a tilde are 
dimensionless quantities and the quantity in the squared brackets comes
from including thermal corrections to the correlation length,
$\xi(T_G)=\xi_0/(2\sqrt{\lambda}) = {1\over {\sqrt{2}\lambda T_G}}$, evaluated
at the Ginzburg temperature $T_G^2\simeq 4\phi_0^2/(1+4\lambda)$, 
the temperature where the
wall network forms in a thermal phase transition \cite{KIBBLE,GGK}.

\begin{figure}
\hspace{2.3in}
\psfig{figure=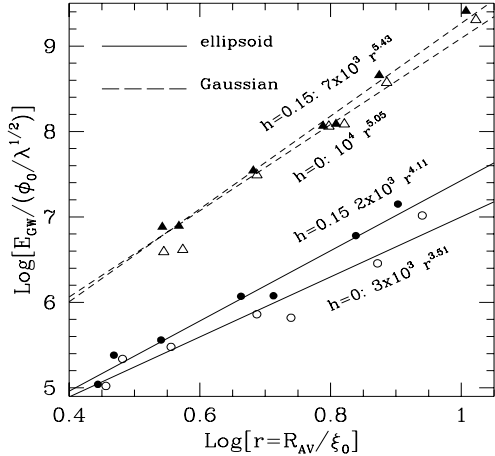,width=2.7in,height=2.5in}
\caption{The integrated output in gravitational radiation generated
during the collapse of the nonspherical domains vs. their initial
average radius. The sample points shown are the same as in Fig. 1.}
\end{figure}

Since the time-scale of collapse is well approximated by the
average size of the domains, the rate of emission is, 

\begin{equation}
\label{g_rate}
\dot E_{\rm GW} \simeq \sqrt{2}{\tilde E}_{\rm GW}^0[2\sqrt{\lambda}]
\left ( {{\phi_0^4}\over {M_{Pl}^2}}\right )\left ({{R_{\rm av}}\over 
{\xi_0}}\right )^{\beta - 1}~.
\end{equation}

The energy density in gravitational radiation for clusters with $r$ and $r+dr$
cells in a time interval between $t$ and $t+dt$ is 

\begin{equation}
\label{rho_noperc}
d\rho_{\rm GW}\simeq \dot E_{\rm GW}~n(r)dr~dt~,
\end{equation}
where $r\simeq (R_{\rm av}/\xi)^3$, and the number density of $r$-clusters can 
be
written as $n(r) = Cr^{-1.5}e^{-r}/V_{\xi}$. Demanding that the
bags occupy a fraction $p_-$ of the horizon volume ($p_-<p_c$), we can fix the
normalization constant as $C\simeq e~p_-$. 

The typical bag lifetime $\tau$ is of order $R_{\rm av}\ll \lambda_H$, 
where $\lambda_H = (45/4\pi^3g_*)^{1/2}M_{Pl}/T^2$ is the horizon length
and $g_*$ is the number of relativistic degrees of freedom 
at temperature $T$. The total output in gravitational radiation is thus,

\begin{equation}
\label{gtot_noperc}
\rho_{\rm GW}\simeq 2\sqrt{2}C\lambda [2\sqrt{\lambda}]^3{\tilde E}_{\rm GW}^0
\phi_0^4\left ({{\phi_0}\over {M_{Pl}}}\right )^2f_{\beta}(r_{\rm max})
~,
\end{equation}
where $f_{\beta}(r_{\rm max})
\equiv \int_1^{r_{\rm max}}r^{\beta/3 -3/2}e^{-r}dr$, is ${\cal O}(1)$ 
for all interesting values of its parameters.

Using that the Universe is radiation-dominated with energy density
$\rho_{\rm rad} = (\pi^2/30)g_*T^4$,
the fraction in energy density from
gravitational waves is, neglecting factors of ${\cal O}(1)$, and using
$T_G\simeq 2\phi_0$,

\begin{equation}
\label{fract_noperc}
\Omega_{\rm GW}\simeq 10^2\lambda [2\sqrt{\lambda}]^3
\left ({{{\tilde E}_{\rm GW}^0}
\over {10^4}}\right )\left ({{100}\over {g_*}}\right )
\left ({{\phi_0}\over {M_{Pl}}}\right )^2~.
\end{equation}

Since the domains will mostly have linear 
dimensions of order $R_{\rm av}\sim \xi$,
the dominant frequency of emission is $\omega \simeq 1/R_{\rm av}\simeq 
\sqrt{2\lambda}[2\sqrt{\lambda}]\phi_0$. Redshifting these quantities 
we obtain,

\begin{equation}
\label{today__Om_noperc}
h^2\Omega_{\rm GW}^0\simeq {{1.7\times 10^{-3}}\over {[2\sqrt{\lambda}]^{-3}}}
\left ({{100}\over {g_*}}\right )^{{3\over 2} }
\left ({{{\tilde E}_{\rm GW}^0}\over {10^4}}\right )\lambda 
\left ({{\phi_0}\over {M_{Pl}}}\right )^2~,
\end{equation}
and,
$
f^0\simeq 1.3\times 10^{10}\sqrt{\lambda}[2\sqrt{\lambda}]
\left ({{100}\over {g_*}}\right )^{{1\over 3}}{\rm Hz}~.
$

Using that the detector strain at wave band given by observational
frequency $f$ for stochastic
gravitational radiation is $h_c(f)\equiv 1.3\times 10^{-18}
[h^2\Omega_{\rm GW}^0]^{1/2}({\rm Hz}/f)$ \cite{THORNE}, we get

\begin{equation}
\label{amp_noperc}
h_c(f)\simeq {{4.1\times 10^{-30}}\over {[2\sqrt{\lambda}]^{-1/2}}}
\left ({{{\tilde E}_{\rm GW}^0}\over {10^4}}
\right )^{{1\over 2}}\left ({{100}\over {g_*}}\right )^{{5\over {12}}}
\left ({{\phi_0}\over {M_{Pl}}}\right )~.
\end{equation}

The no-percolation case
is characterized by a very high frequency spectrum and small amplitudes, 
beyond the presently projected sensitivity
of ground-based interferometers.
The situation is quite different for the
percolating case.

When both vacua percolate, they are separated by a convoluted domain wall,
with initial average curvature $R_{\rm av}\simeq \xi$. The tension force 
dominates the wall dynamics for a while, until
the vacuum pressure starts accelerating the walls against each other, eventually
causing the formation of large, unstable domains. 
In order for this to happen, the
asymmetry must satisfy 
$h < {{3\sqrt{2}}\over 2}\lambda[2\sqrt{\lambda}]^3\tilde E_0^0$, while,
for the walls to disappear as their average curvature reaches  cosmologically
significant scales ($R_{\rm av}\lesssim \lambda_H$),
$h\gtrsim (\lambda )^{1/2+(\gamma-2)}(\sqrt{g_*}\phi_0/M_{Pl})^{3-\gamma}
\tilde E_0^0$. 
To these bounds we add the
percolation constraint, $p_-\geq 0.31$, which becomes, $h\leq 0.15\lambda$ 
\cite{GGK}.

Writing the average radius of 
the domains as $R_{\rm av} \equiv a\lambda_H$, 
the energy density in gravitational waves from collapsing
domains is, ($a$ is not the same as the ellipsoidal axis)
$
\rho_{\rm GW} \simeq {{\dot E_{GW}}\over {a^2\lambda_H^2}}$, or
with $\kappa \equiv 45/32\pi^3$, 

\begin{equation}
\rho_{\rm GW} \simeq  
{{2\tilde E_{\rm GW}^0\phi_0^4
a^{\beta-3}\lambda^{(\beta-1)/2}}\over 
{\kappa^{(3-\beta)/2} g_*^{(\beta-3)/2}}}
\left ({{\phi_0}\over {M_{Pl}}}\right )^{5-\beta}~.
\end{equation}

The typical frequency will be, $f_{\rm av}\simeq [2\pi a\lambda_H]^{-1}$,
which redshifts to 

\begin{equation}
\label{today_F_perc}
f^0 \simeq 6.1\times 10^{11}a^{-1}\left ({{g_*}\over {100}}\right )^{
{1\over 6}}
\left ({{\phi_0}\over {M_{Pl}}}\right ){\rm Hz}~,
\end{equation}
while the fraction of the energy density in gravitational waves today is

\begin{equation}
\label{today_Omega_perc}
h^2\Omega_{\rm GW}^0\simeq {{6.7 \times 10^{-\beta+1}
\left ({{{\tilde E}_{\rm GW}^0}\over
{10^4}}\right )
\left ({{100}\over {g_*}}\right )^{{{\beta - {1\over 3}}\over 2}}}\over 
{a^{3-\beta}\lambda^{(1-\beta)/2}\kappa^{-\beta/2} }}
\left ({{\phi_0}\over {M_{Pl}}}\right )^{5-\beta}~.
\end{equation}

The detector strain in the wave band $f$ is given by,

\begin{equation}
h_c(f)\simeq {{1.7\times 10^{-29-{{\beta}\over 2}}
\left ({{{\tilde E}_{\rm GW}^0}\over{10^4}}\right )^{{1\over 2}}
\left ({{100}\over {g_*}}\right )^{{{\beta + {1\over 3}}\over 4}}}\over 
{a^{(1-\beta)/2}\lambda^{(1-\beta)/4}\kappa^{-\beta/4}}}
\left ({{\phi_0}\over {M_{Pl}}}\right )^{{{3 - \beta}\over 2}}~.
\end{equation}

We have summarized our results in Fig. 3, where 
the detector strain vs. 
frequency are plotted for several
values of $\beta$ and compared with the LIGO (initial and advanced)
and LISA sensitivities. The solid squares locate an electroweak
phase transition at $T=200$ GeV. The solid triangles, a hypothetical transition
at $T=10^9$ GeV. Continuous curves are for ellipsoids while dashed curves are
for Gaussians. The dot-dashed curve is for a Gaussian with $a=10^{-2}$.
LIGO is mostly sensitive to collapsing
walls at $\phi_0\sim 10^9$ GeV or so, while LISA can probe the
electroweak phase transition for a comfortable range of parameters. The 
spectrum from collapsing bags, being very pronounced at a given frequency,
is quite different from the mostly flat spectra of cosmic strings and
inflationary models \cite{ALLEN}. A more detailed analysis will be presented 
in a forthcoming publication. 

\begin{figure}
\hspace{2.3in}
\psfig{figure=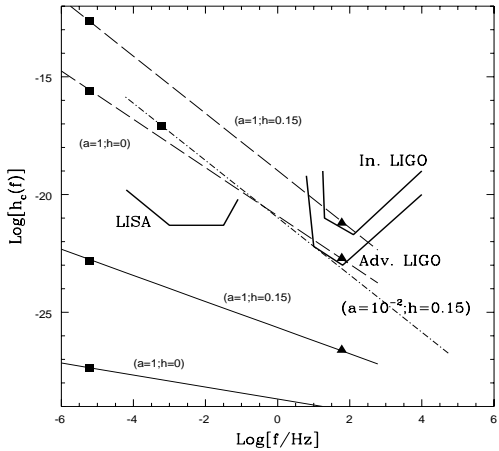,width=2.7in,height=2.5in}
\caption{Predicted detector strain vs. dominant frequency band from
collapsing nonspherical domains. We also display the sensitivities of the LIGO
and LISA detectors. The solid cubes are for an electroweak
transition at $T=200$ GeV and the solid triangle for a hypothetical
transition at $\phi_0=10^9$ GeV.}
\end{figure}

It is clear that the stochastic background of gravitational
waves left behind from collapsing domain walls can be detectable for several
situations of interest, illustrating the potential impact of 
gravitational wave astronomy on applications of
particle physics to the early Universe.

The authors were partially supported 
by the NSF through a Presidential
Faculty Fellows Award no. PHY-9453431 and by a NASA
grant no. NAG5-6613.

\end{document}